\documentclass[10pt]{article}

\usepackage{amssymb}
\usepackage{bm}% bold math
\usepackage{verbatim}       % Defines \begin{comment}, \end{comment}
\usepackage{amsthm}
\usepackage{amssymb}

\begin{document}

\title{Globally hyperbolic spacetimes can be defined as
``causal'' instead of ``strongly causal''}

\author{Antonio N. Bernal and Miguel S\'anchez
\thanks{The careful reading by J.M.M. Senovilla is
warmly acknowledged. These problems were discussed in the
interdisciplinary meeting ``Hyperbolic operators and
quantization'' (Erwin Schr\"odinger Institute, Vienna, November
2005) organized by C. B\"ahr, N. Ginoux and F. Pf\"affle, within
the ESI-semester ``Geometry of Pseudo-Riemannian Manifolds with
Applications in Physics'' organised by D. Alekseevsky, H. Baum,
and J. Konderak. The author acknowledges their support. This work
is partially economically supported by a MEC-FEDER
Grant No. %\newline
 MTM2004-04934-C04-01.}
\\Departamento de Geometr\'{\i}a y Topolog\'{\i}a \\
Facultad de Ciencias, Universidad de Granada \\
 18071-Granada, Spain.\\
E-mail: {\ttfamily  sanchezm@ugr.es}}

\date{  }
\textwidth 6.5 in \textheight 8.0 in \topmargin 0 in
\oddsidemargin 0.7 in \evensidemargin 0.7 in

\makeatletter

\def\theequation{\thesection.\arabic{equation}}
\@addtoreset{equation}{section}

\newtheorem{defn}{Definition}
\def\thedefn{\thesection.\arabic{defn}}
\@addtoreset{defn}{section}

\newtheorem{teo}[defn]{Theorem}
\newtheorem{eje}[defn]{Example}
\newtheorem{lem}[defn]{Lemma}
\newtheorem{rem}[defn]{Remark}
\newtheorem{cor}[defn]{Corollary}
\newtheorem{pro}[defn]{Proposition}

\makeatother \font\ddpp=msbm10  at 10 truept
\def\R{\hbox{\ddpp R}}
\def\C{\hbox{\ddpp C}}
\def\L{\hbox{\ddpp L}}
\def\S{\hbox{\ddpp S}}
\def\Z{\hbox{\ddpp Z}}
\def\N{\hbox{\ddpp N}}

\newcommand{\D}{{\cal D}}
\newcommand{\M}{{\cal M}}
\newcommand{\Mo}{{\cal M}_0}
\newcommand{\la}{\Lambda}
\newcommand{\dimo}{{\bf Proof: }}
\newcommand{\inte}{\int_{0}^{1}}
\newcommand{\gam}{\gamma}
\newcommand{\eps}{\epsilon}
\newcommand{\<}{\langle}
\renewcommand{\>}{\rangle}
\newcommand{\Om}{\Omega^1}
\renewcommand{\(}{\left(}
\renewcommand{\)}{\right)}
\newcommand{\om}{\omega}
\newcommand{\me}{\frac{1}{2}}
\newcommand{\Mt}{\widetilde{\M}}
\newcommand{\cat}{{\mathop{\rm cat}\nolimits}}

\newcommand{\cvd}{\ \rule{0.5em}{0.5em}}

\newcommand{\be}{\begin{equation}}
\newcommand{\ee}{\end{equation}}
\newcommand{\noi}{\noindent}
\newcommand{\ben}{\begin{enumerate}}
\newcommand{\een}{\end{enumerate}}
\newcommand{\bit}{\begin{itemize}}
\newcommand{\eit}{\end{itemize}}
\newcommand{\edoc}{\end{document}}

\newcommand{\p}{\partial}
\newcommand{\eu}{{\rm e}}
\newcommand{\dd}{{\rm d}}
%\renewcommand{\thesection}{\arabic{section}}
%%%%%%%%%%%%%%%%%%%%%%%%%%%%%%%%%%%%%%%%%%%%%%%%%%%%%%%%%%%%%%%%%%%%%
\newcommand{\bd}{\begin{definition}}                %inizia definizione
\newcommand{\ed}{\end{definition}}                  %fine definizione
\newcommand{\bc}{\begin{corollary}}                 %inizia corollario
\newcommand{\ec}{\end{corollary}}                   %fine corollario
\newcommand{\bl}{\begin{lemma}}                     %inizia lemma
\newcommand{\el}{\end{lemma}}                       %fine lemma
\newcommand{\bp}{\begin{proposition}}            %inizia proposizione
\newcommand{\ep}{\end{proposition}}                %fine proposizione
\newcommand{\bere}{\begin{remark}}                  %inizia osservazione
\newcommand{\ere}{\end{remark}}                     %fine oservazione

\newcommand{\bt}{\begin{theorem}}
\newcommand{\et}{\end{theorem}}

\newtheorem{theorem}{Theorem}[section]
\newtheorem{corollary}[theorem]{Corollary}
\newtheorem{lemma}[theorem]{Lemma}
\newtheorem{proposition}[theorem]{Proposition}
\newtheorem{definition}[theorem]{Definition}

\newtheorem{remark}[theorem]{Remark}
\newtheorem{example}[theorem]{Example}

%%%%%%%%%%%%%%%%%%%%%%%%%%%%%%%%%%%%%%%%%%%%%%%%%%%%%%%%%%%%%%%%%%%%%%

\hyphenation{Lo-rent-zian}

\maketitle \vspace*{-6mm}

\begin{abstract}
The classical definition of {\em global hyperbolicity} for a
spacetime $(M,g)$ comprises two conditions: (A) compactness of the
diamonds $J^+(p)\cap J^-(q)$, and (B) strong causality. Here we
show that condition (B) can be replaced just by causality. In
fact, we show first that the classical  definition of causal
simplicity (which impose  to be distinguishing, apart from the
closedness of $J^+(p)$, $J^-(q)$) can be weakened in  causal
instead of distinguishing. So, the full consistency of the causal
ladder (recently proved by the authors in a
definitive way) yields directly the result.
\end{abstract}
 \vspace{2mm}



%\hyphenation{eve-ry} %\newpage

\section{Introduction}

Global hyperbolicity is the most important condition on Causality,
which lies at the top of the so-called causal hierarchy of
spacetimes and is involved in problems as Cosmic Censorship,
predictability etc. There are different alternative definitions of
what global hyperbolicity means, but perhaps the most standard one
is the following. A spacetime $(M,g)$ is said {\em globally
hyperbolic} if and only if it satisfies two conditions: (A)
compactness of $J^+(p)\cap J^-(q)$ for all $p,q\in M$ (i.e. no
``naked'' singularity can exist) and (B) strong causality (no
``almost closed'' causal curve exists).

In this note, we stress the possibility to simplify the definition
of global hyperbolicity,  by weakening the requirement of {\em
strong causality} into {\em causality} (i.e. no closed causal
curve exist), which is both, mathematically simpler and physically
clearer. More in depth, we will see:

\begin{enumerate} \item The usual
definition of causal simplicity for a spacetime $(M,g)$ can be
simplified, just by imposing:

(a) closedness of $J^\pm (p)$ for all $p$, and

(b) causality.

\item In any spacetime, the condition:

(A) compactness of $J^+(p)\cap J^-(q)$ for all $p,q\in M$

always implies previous condition (a).

\item Thus, any  spacetime which satisfies (A) and (b) is causally
simple. As this condition implies strong causality, {\em any
spacetime which satisfies (A) and (b) is strongly causal and,
then, globally hyperbolic}. \een

About this last point, it is worth pointing out  that there were
some problems on the full consistency of the causal hierarchy of
spacetime, which has been recently solved (\cite{bernal03, bernal04}, 
see also the reviews \cite{Sa,ms}). Concretely (see Remark \ref{r} below), the so-called ``folk
questions'' on smoothability posed a question on the equivalence
between two alternative definitions of {\em stable causality},
i.e., the level in the standard hierarchy of causality which is
immediately more
 restrictive than strong causality. Nevertheless, after the solution of
 these folk
 problems,  no doubt can exist in the use of the full consistency
 and its classical
 implications (a detailed study is done in the survey \cite{ms}).

 In the next Sections \ref{s1}, \ref{s2} we prove and discuss these
 simplifications of causal simplicity and global hyperbolicity,
 respectively. All the notation and concepts are standard, and can
 be found, for example, in
 \cite{beem96, hawking73, oneill83, penrose72, Se, wald84}.

\section{The optimal definitions of causal simplicity}\label{s1}

Let us start with the following result:

\begin{pro} \label{p2}
Assume that $(M,g)$ satisfies: \bit \item[($a$)] $J^+(p)$ and
$J^-(p)$ are closed for all $p,q \in M$. \eit Then the following
two conditions are equivalent: \bit \item[($b$1)] $(M,g)$ is
causal,
 \item[($b$2)] $(M,g)$ is distinguishing, i.e., if $p\neq q$ then $I^+(p)\neq I^+(q)$ and
 $I^-(p)\neq I^-(q)$.\eit
\end{pro}

\begin{proof}  Let us prove that, under
($a$), hypothesis ($b$1) imply ($b$2) (the converse holds always
and is well-known). Otherwise, if $p\neq q$ and, say
$I^+(p)=I^+(q)$, choose any sequence $\{q_n\}\rightarrow q$, with
$q\ll q_n$. Then, $q\in J^+(p)$ because $q\in \bar{I}^{+}(q) =
\bar{I}^+(p) = \bar{J}^+(p) = J^+(p)$ (the first equality holds
because the distinguishing property fails for $p$ and $q$, the
second one holds always, and the last one by hypothesis ($a$)).
Analogously, $p\in J^+(q)$, i.e., $p<q<p$ and the spacetime is not
causal, a contradiction. \end{proof}

\begin{remark} {\em The classical definition of causal simplicity \cite[p.
65]{beem96}, \cite[Dfn. 2.17]{Se}
 assumes\footnote{In \cite[p.
188]{hawking73}, there is a small gap, because no condition type
($b$) is assumed (in this case, any totally vicious spacetime
would be causally simple). } ($a$) and ($b$2). But ($b$2) is
always more restrictive than ($b$1) and, thus,
 it is natural to choose a definition of causal simplicity with
minimum hypotheses, that is, {\em  a causally simple spacetime is
the one which satisfies ($a$) and ($b$1)}. Even more, it is easy
to  show that the requirement ($b$1) cannot be weakened in $(M,g)$
chronological (see the cylinder $C$ constructed in item 2 below
Remark \ref{r}). }
\end{remark}

\section{The optimal definition of global hyperbolicity}\label{s2}

The following result is rather standard in Causality, in order to
prove that a globally hyperbolic spacetime is causally simple (see
for example \cite[Prop. 2.26]{Se}):

\begin{lem} \label{l1}
Assume that a spacetime $(M,g)$ satisfies: \bit \item[($A$)]
$J^+(p)\cap J^-(q)$ is compact for all $p,q \in M$. \eit Then it
also satisfies:

\bit \item[($a$)] $J^+(p)$ and $J^-(p)$ are closed for all $p,q
\in M$. \eit
\end{lem}

\smallskip

\begin{proof} Assume that $J^+(p)$ is not closed and
choose $r\in \bar{J}^+(p)\backslash J^+(p)$ and $q\in I^+(r)$.
Take a sequence $\{r_n\}\rightarrow r$ with $r_n\in I^+(p)$ for
all $n$, and notice that $r_n \ll q$ up to a finite number of $n$
(as $I^-(q)\ni r$ is open). Thus, $\{r_n\}_n \subset J^+(p)\cap
J^-(q)$, which is compact, but converges to the point $r$, which
does not lie in this  subset, a contradiction. \end{proof}

 Now, the next result yields the simplification of global hyperbolicity; it becomes a straightforward consequence of previous ones and the full
 consistency of the causal ladder of causality.

\begin{teo} \label{p1}
Assume that $(M,g)$ satisfies: \bit \item[(A)] $J^+(p)\cap J^-(q)$
is compact for all $p,q \in M$. \eit Then the following two
conditions are equivalent: \bit \item[(B1)] $(M,g)$ is causal,
i.e., there are no closed causal curves. \item[(B2)] $(M,g)$ is
strongly causal, i.e., for any $p\in M$, given any neighborhood
$U$ of $p$ there exists a neighborhood $V\subset U$, $p \in V$,
such that any future-directed (and hence also any past-directed)
causal curve $\gamma: [a,b]\rightarrow M$ with endpoints at $V$ is
entirely contained in $U$.\eit
\end{teo}

\smallskip

\begin{proof} By Lemma \ref{l1} (A) plus (B1) imply conditions ($a$) plus ($b$1) in Proposition \ref{p2} i.e.,
causal simplicity. This is a more restrictive level of the causal
ladder than strong causality (a standard reference is \cite[p.
73]{beem96}, but see discussion below), and thus, the result
follows.
 \end{proof}

\begin{rem} \label{r} {\em Notice that, in  Theorem \ref{p1}, the proof of strong causality follows indirectly,
as a consequence of causal simplicity. Taking into account the
known properties of the hierarchy of spacetimes, if $(M,g)$ is
causally simple  then it is {\em causally continuous} and, thus,
the volume functions $t^\pm(p)= m(J^\pm(p))$ (defined for any {\em
admissible measure} $m$), are -continuous- time functions. In
particular, $(M,g)$ satisfies one of the two
definitions\footnote{The other essentially different definition is
a spacetime which remains causal after opening a bit all the
causal cones.}
  of {\em stable causality}, and it is strongly causal.
The full consistency between these two alternative definitions
(and, thus, of the whole causal ladder) has been obtained recently
\cite{bernal04, bernal03}; we refer to \cite{Sa, ms} for details
and subtleties. }\end{rem}

\noi Now, we emphasize the following items:

\ben \item The classical definition of global hyperbolicity
(\cite[p. 65]{beem96}, \cite[p. 206]{hawking73}, \cite[p.
412]{oneill83}, \cite[p. 48]{penrose72}, \cite[Dfn. 2.17]{Se},
\cite[p. 209]{wald84}) impose (A) and (B2). Nevertheless,
(B2)$\Rightarrow $ (B1) always trivially and, thus, it is natural
to choose a definition of global hyperbolicity with minimum
hypotheses, that is, {\em  a globally hyperbolic spacetime is the
one which satisfies (A) and (B1)}.

\item The requirement (B1), i.e.,  the spacetime is causal, cannot
be weakened in $(M,g)$ chronological. In fact, the cylinder $C$
obtained as the quotient from Lorentz-Minkowski spacetime in null
coordinates $(\R^2, g= 2dudv )$ by the group generated by the
translation $(u,v)\rightarrow (u+1,v)$ is chronological, satisfies
(A) but it is not causal.

\item An alternative hypothesis to condition (A) is:

(A') For each $p,q \in M$, the space of future-directed causal
 curves (continuous, and up to a strictly increasing reparameterization) which connect $p$ with
 $q$, $C(p,q)$, is compact in the $C^0$ topology.

So, conditions (A') and (B1) are also equivalent to global
hyperbolicity. Even more, the use of the $C^0$ topology makes
sense only when causality holds (otherwise, if $\gamma$ is a
closed causal curve through $p$, giving more and more rounds one
would obtain non-equivalent curves; clearly, it would be natural
to consider a topology such that $C(p,p)$ is not compact). So,
essentially global hyperbolicity reduces to (A').

\item Due to a classical theorem by Geroch \cite{geroch70},
globally hyperbolic spacetimes are alternatively defined as the
ones which admit a Cauchy hypersurface. Even more, the recent
solution of the ``folk'' questions on smoothability
\cite{bernal04}, yield further orthogonal metric structures for
globally hyperbolic spacetimes. Nevertheless, these alternative
possibilities are formulated under  hypotheses different to (A)
and (B1), (B2) and, in principle, they are not affected by Theorem
\ref{p1} (even though some relations are still possible; for
example, recall that  if $S$ is an achronal subset then $D(S)$ is
strongly causal). \een

%\newpage

\end{document}